\documentclass[prl,superscriptaddress,twocolumn,showpacs,preprintnumbers,amsmath,amssymb]{revtex4}

\usepackage{graphicx}
\usepackage{bm}

\def\etal{{\it et~al.}}

\def\cm{\ {\rm cm^{-1}}}

\def\s1{\sigma_1(\omega)}
\def\er{\epsilon_1}
\def\ei{\epsilon_2}

\def\cmtwo{\ {\rm cm}^{-2}}
\def\cmthree{\ {\rm cm}^{-3}}

\begin{document}

\preprint{???}

\title{\boldmath Dynamical response and confinement of the electrons at the LaAlO$_3$/SrTiO$_3$ interface \unboldmath}

\author{A. Dubroka}
\author{M. R\"{o}ssle}
\author{K. W. Kim}
\author{V. K. Malik}
\author{L. Schultz}
 \affiliation{University of Fribourg, Department of Physics and Fribourg Center for Nanomaterials,
Chemin du Mus\'{e}e 3, CH-1700 Fribourg, Switzerland}
\author{S. Thiel}
\author{C.~W.~Schneider}
\altaffiliation{PSI, Materials Group, 5232, Villigen, Switzerland}
\author{J. Mannhart}
\affiliation{Experimental Physics VI, Center for Electronic Correlations
and Magnetism, Institute of Physics, University of Augsburg,
D-86135 Augsburg, Germany.}
\author{G.~Herranz}
\altaffiliation{Institut de Ci\`encia de Materials de Barcelona, ICMAB-CSIC Campus de la UAB, Bellaterra 08193, Spain}
\author{O.~Copie}
\author{M. Bibes}
\author{A. Barth\'el\'emy}
\affiliation{Unit\'e Mixte de Physique CNRS/Thales associ\'ee \`a l'Universit\'e Paris-Sud, 
Campus de l'Ecole Polytechnique, 1 Avenue A. Fresnel, 91767 Palaiseau, France}
\author{C. Bernhard}%
 \email{christian.bernhard@unifr.ch}
\affiliation{University of Fribourg, Department of Physics and Fribourg Center for Nanomaterials,
Chemin du Mus\'{e}e 3, CH-1700 Fribourg, Switzerland}

\date{\today}

\begin{abstract}
With infrared ellipsometry and transport measurements we investigated the electrons at the interface between LaAlO$_3$ and SrTiO$_3$. We obtained a sheet carrier density of $N_{\rm s}\approx5-9\times 10^{13}$ cm$^{-2}$, an effective mass of $m^*\approx 3m_e$, and a strongly frequency dependent mobility. The latter are similar as in bulk SrTi$_{1-x}$Nb$_x$O$_3$ and therefore suggestive of polaronic correlations of the confined carriers. We also determined the vertical density profile which has a strongly asymmetric shape with a rapid initial decay over the first 2~nm and a pronounced tail that extends to about 11~nm. 
\end{abstract}

\maketitle

It has been demonstrated that highly mobile charge carriers can develop at the interface between a SrTiO$_3$ (STO) substrate and a thin LaAlO$_3$ (LAO) layer that is grown on top~\cite{Ohtomo,Nakagawa,Thiel,Herranz,Kalabukhov,Siemons,Basletic,Reyren,Brinkman,Breitschaft}. 
This came as a surprise since both materials are insulators with $\Delta_{\rm gap}^{\rm STO}=3.2$~eV and $\Delta_{\rm gap}^{\rm LAO}=5.6$~eV, respectively. It has been suggested that the carriers originate from an electronic reconstruction across the heteropolar interface which gives rise to a transfer of 1/2 {\it e} charge per unit cell yielding a sheet carrier density of 
$N_{\rm s}=3.3\times 10^{14}\cmtwo$. Supporting this so-called polarization catastrophe scenario, transport measurements showed that a critical thickness of the LAO layer of 4 unit cells and thus a minimal electrostatic potential is required for the conducting layer to develop~\cite{Thiel}, albeit the measured
 $N_{\rm s}$ is only $2-6\times 10^{13}\cmtwo$. Alternatively, it has been argued that the carriers are induced by chemical doping due to oxygen vacancies~\cite{Kalabukhov} or interfacial mixing of La and Sr \cite{Willmott}. While this question remains unsettled, it was recently shown that the carrier density can be largely varied by applying a gate voltage, thus allowing for a reproducible switching between insulating, metallic and even superconducting states~\cite{Thiel,Caviglia}. These exciting experimental developments inspired large theoretical efforts~\cite{Pentcheva,Janicka} and raised hopes that these oxides may be useful for new electronic devices~\cite{Cen2} and for studying electric field induced quantum phase transitions. Accordingly, the investigation of the fundamental properties of these confined charge carriers is a subject of utmost importance. Here we present infrared (IR) ellipsometry data which provide direct information about their dynamical properties and their confinement.
 
The LAO/STO heterostructures were grown by pulsed laser deposition (PLD) with in-situ monitoring of the LAO layer thickness by reflection high energy electron diffraction (RHEED). Two samples were grown in Augsburg at an oxygen pressure of $6\times10^{-5}$~mbar at 770$^\circ$C. They were subsequently cooled to 300~K in 400~mbar of O$_2$ with a 1h oxidation step at 600$^\circ$C to remove any oxygen vacancies in the STO substrate which could induce a bulk metallic state~\cite{Siemons}. The LAO layer thickness of 3 and 5 unit cells (LS-3 and LS-5) was chosen to lie below and above the critical value where a conducting interface layer was shown to develop~\cite{Thiel}. The samples LS-50 and LS-50-ov were grown in Palaiseau as described in Ref.~\cite{Herranz} at oxygen pressures of $10^{-4}$ and $10^{-6}$ mbar so the bulk of the STO substrate is insulating and conducting, respectively~\cite{Siemons}.

Ellipsometry enables accurate and direct measurements of the real- and imaginary parts of the pseudo-dielectric function, $\er$ and $\ei$. The far- to mid-infrared (FIR-MIR) measurements were performed with a home-built setup attached to a Bruker 113V Fast-Fourier spectrometer~\cite{Bernhard}. The angle of incidence of the light was 75 degree; a ZnSe compensator was used above 700$\cm$. Photodoping~\cite{Kozuka} was avoided by shielding the sample against visible and UV light.

\begin{figure*}
\vspace{-1.2cm}
\hspace*{-0.8cm}\includegraphics[width=19cm]{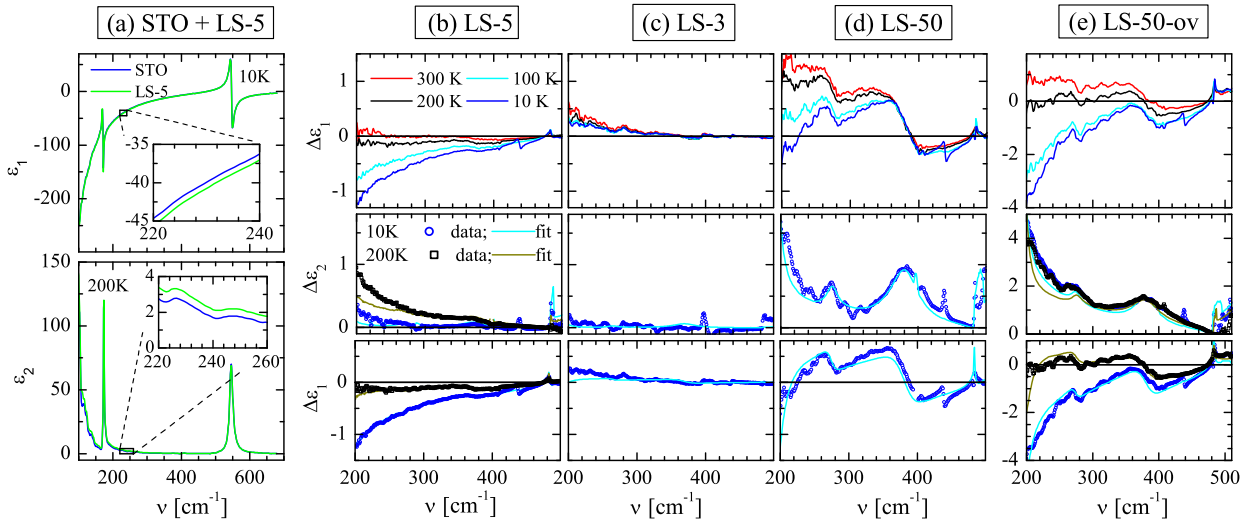}
\vspace{-1.2cm}
\caption{\label{FIR} (a) Spectra of $\er$ (10~K) and $\ei$ (200~K) for LS-5 (green lines) and bare SrTiO$_3$ (blue lines). Insets: Magnification of the small, yet significant differences. Corresponding spectra of $\Delta\epsilon_{1,2}=\epsilon_{1,2}-\epsilon_{1,2}({\rm STO})$ for samples LS-5 (b), LS-3 (c), LS-50 (d), and LS-50-ov (e) showing the signatures of the Drude-response. Top panels show the T-dependence of $\Delta\epsilon_{1}$. Middle and bottom panels show $\Delta\epsilon_{1}$  and $\Delta\epsilon_{2}$ as obtained from experiment (symbols) and model calculations (lines), respectively.}
\end{figure*}

Figure~1 shows our ellipsometry data which reveal the Drude response of the mobile electrons at the LAO/STO interface. Figure 1(a) displays the spectra of $\er$ and $\ei$ for LS-5 and bare STO. They almost coincide and are dominated by the STO phonons near 545, 175 and the strong so-called soft mode below 100$\cm$. Nevertheless, as shown in the insets, some small, yet significant and reproducible differences appear upon magnification. Figure~1(b) shows the corresponding difference spectra of $\Delta\epsilon_{1,2}=\epsilon_{1,2}($LS-5$)-\epsilon_{1,2}({\rm STO})$. Here the  inductive response of the charge carriers is evident from the temperature (T) and frequency dependent decrease of $\Delta\er$ (upper panel). The data are well reproduced (middle and bottom panel) with a model (solid lines) that contains a conducting layer at the LAO/STO interface with a block-like profile of the carrier concentration with thickness $d=10$~nm;  for details see Ref.~\cite{EPAPS}. The model explains even the minor features of the spectra~\cite{EPAPS}. Figures~1(c) and (d) show our data on LS-3 and LS-50 which confirm that we are probing the Drude response of interfacial charge carriers. For LS-3 there is no clear indication of an inductive decrease of $\Delta\er$ and thus of a Drude response. For LS-50, the magnitude of the inductive response is similar as in LS-5. These trends agree with a previous report that the conducting layer develops only above a LAO thickness of 4 unit cells~\cite{Thiel}. 
Finally, Fig.~1(e) shows our data for LS-50-ov where even the bulk of the STO substrate is conducting~\cite{Basletic}. Its inductive response is indeed significantly stronger than in LS-50 and LS-5 (note the enlarged vertical scale).
 
The result of a quantitative analysis and the comparison with the transport data are detailed in Fig.~2. The dielectric response is well approximated by the volumetric average of $\epsilon$ of the individual layers since $\lambda_{\rm IR}\gg p_{\rm IR}\gg d$, where $\lambda_{\rm IR}$ is the wave length and $p_{\rm IR}$ the penetration depth of the IR light. From the strength of the Drude-response we thus derive $d\omega_{\rm pl}^2=N_{\rm s}^{\rm opt}e^2/\epsilon_0m^*$, where $\omega_{\rm pl}$, $m^*$, $e$, and $\epsilon_0$ are the plasma frequency, effective carrier mass, electric charge and dielectric constant, respectively. Furthermore, we obtain the scattering rate, $\gamma$, or the electron mobility, 
$\mu^{\rm opt}=e/(m^*2\pi c\gamma)$. In addition, $m^*$ can be deduced under the condition that the IR and transport measurements are probing the same laterally homogeneous electron system and thus yield the same values for $N_{\rm s}^{\rm opt}$ and $N_{\rm s}^{\rm tr}$.  Figures~2(a) and~(b) show that a good agreement between the IR and transport data of LS-5 is obtained for $T\geq100$~K with $m^*=3.2 m_e$; this value is used in further analysis throughout this paper. Notably, this value is similar as in bulk SrTi$_{1-x}$Nb$_x$O$_3$~\cite{Mechelen} where the electrons reveal polaronic correlations. In the latter $m^*$ is found to gradually decrease at low T. The gradual increase of $N_{\rm s}^{\rm opt}$ as shown in Fig. 2(a) (assuming a constant $m^*=3.2 m_e$) thus may instead be the signature of a corresponding decrease of $m^*$. The sudden decrease of $N_{\rm s}^{\rm tr}$ below 100~K likely has a different origin, possibly due to a weak localization which affects the dc response but hardly the IR one. The reason might be a tilting of the heterostructure below a structural transition of the STO substrate which has been observed on a different kind of oxide heterostructure~\cite{Hoppler08}.

In comparing the different samples, we find that LS-5 and LS-50 have similar values of 
$N_{\rm s}^{\rm opt}\approx 9\times10^{13}\cmtwo$ at 10~K. This is despite of a tenfold difference in the thickness of the LAO layer and the fact that the samples were grown in different conditions and growth chambers. A significantly enhanced Drude response is only observed in LS-50-ov where even the bulk of the STO substrate is conducting. Nevertheless, due to the probe depth of $p_{\rm IR}\approx 1\ \mu$m, our IR experiment captures only a small fraction of these bulk charge carriers yielding 
$N_{\rm s}^{\rm opt}\approx 4\times10^{14}\cmtwo\ll N_{\rm s}^{\rm tr}\approx 10^{16}\cmtwo$. 
The difference in probe depth can also explain the discrepancy between the low-T values of $\mu^{\rm opt}\approx 20\ {\rm cm^2/Vs}$ and
 $\mu^{\rm tr}\approx10^4\ {\rm cm^2/Vs}$ of LS-50-ov~\cite{Herranz} if the carriers near the interface are more strongly scattered on defects than the ones in the bulk. 
A strong scattering of the confined carriers on defects, possibly related to the growth conditions and additional strain due to a thicker LAO layer, is also suggested by the difference between $\mu^{\rm opt}\approx\mu^{\rm tr}\approx 700\ {\rm cm^2/Vs}$ at 10~K in LS-5 and $\mu^{\rm opt}\approx 20\ {\rm cm^2/Vs}$ and $\mu^{\rm tr}\approx 10\ {\rm cm^2/Vs}$ at 10~K in LS-50~\cite{Herranz}.

\begin{figure}
\vspace*{-0.6cm}
\hspace*{-0.8cm}
\includegraphics[width=10cm]{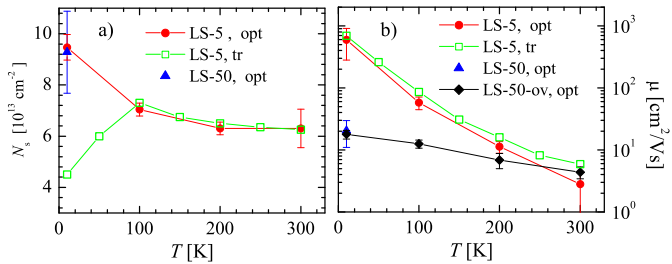}
\vspace*{-1.2cm}
\caption{\label{Mobility}Comparison of the sheet carrier density $N_{\rm s}$ (a) and mobility $\mu$ (b) as deduced from IR (full symbols) and transport (open symbols) measurements. The data for LS-5 are scaled by adjusting the effective mass to $m^*=3.2 m_e$. The same value is used for LS-50 and LS-50-ov. Error bars represent one standard deviation.}
\end{figure}

\begin{figure*}
\vspace{-0.5cm}
\includegraphics[width=18cm]{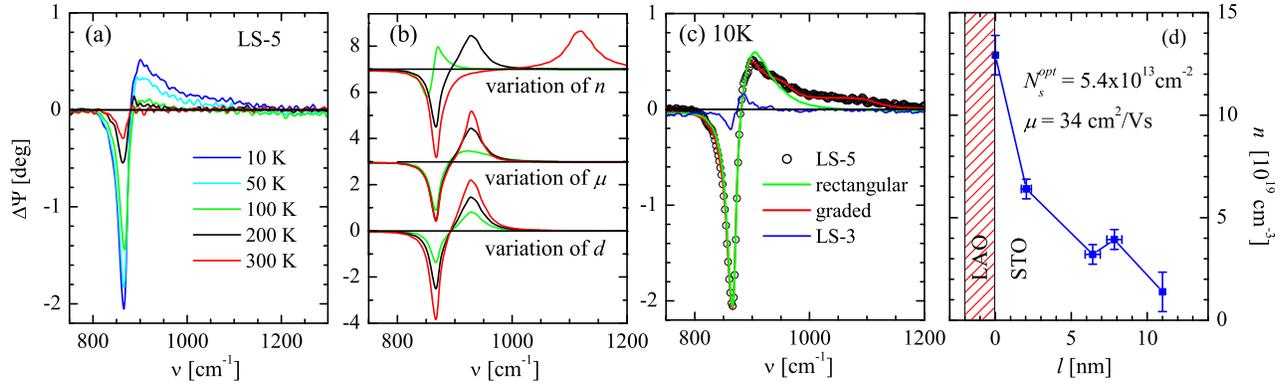}
\vspace{-0.5cm}
\caption{\label{Berreman} (a) Difference spectra of the ellipsometric angle, $\Delta\Psi=\Psi({\rm LS}$-$5)-\Psi({\rm STO})$, showing the Berreman mode in the vicinity of the highest LO phonon of STO. (b) 
Simulations of $\Delta\Psi$ with a block-like density profile of the conducting layer. Shown from top to bottom (shifted for clarity) are the changes with respect to the black line ($d=11~{\rm nm},\ n=5\times10^{13}\cmtwo,\ \mu=33\ {\rm cm^2/Vs}$) upon variation of $n$ (green, $2.2\times10^{13}$, red $13\times10^{13}$), of $\mu$ (green $10\ {\rm cm^2/Vs}$, red $80\ {\rm cm^2/Vs}$), and $d$ (green 6~nm,  red 17~nm). 
(c) Comparison between the data of LS-5 at 10~K (open circles) and calculations assuming a rectangular (green line) and a graded profile (red lines) of the conducting layer. The corresponding data for LS-3 are shown as blue line. (d) Depth profile of $n$ as obtained from the fit (red line) in (c). Error bars represent one standard deviation. }
\end{figure*}

Next we present our data and analysis of a Berreman mode~\cite{Berreman} which provide insight into the vertical density profile of the mobile carriers~\cite{Humlicek}. Berreman modes arise from a dynamical charge accumulation at interfaces of heterostructures that is driven by the normal component of the polarized light. Accordingly, they gives rise to IR-active dipoles and corresponding resonances near the frequencies of the longitudinal optical (LO) modes of the constituent materials~\cite{Sihvola}. These features show up in the reflection coefficient, $r_{\rm p}$, for parallel (p) polarization with respect to the plane of incidence and are absent in the coefficient, $r_{\rm s}$, for perpendicular (s) polarization. Accordingly, as shown in Fig.~3, the signatures of a Berreman mode are well represented in terms of the ellipsometric angle, $\Psi={\rm arctan}\,(|r_{\rm p}/r_{\rm s}|)$. Figure 3(a) shows the T-dependence of the difference spectrum, $\Delta\Psi$, between LS-5 and pure STO. It contains a feature with a sharp minimum around $865\cm$ and a broader maximum near $900\cm$ that becomes pronounced at low~T. The latter corresponds to the structure in $r_{\rm p}$ near $\omega_{\rm LO}$ of the conducting layer. 
The former arises from the polarization dependence of the reflection coefficient near 
$\epsilon_1({\rm STO})=1$ where the reflection nearly vanishes. 
Figure~3(b) displays simulations which show that these features in $\Delta\Psi$ undergo characteristic changes as a function of $n$, $\mu$, and $d$. For the minimum, the variation of these parameters mostly affects the magnitude but hardly the shape and position. For the maximum, the center frequency strongly increases as a function of $n$ while it hardly changes with $\mu$ and $d$. The upward shift of the maximum with respect to $\omega_{\rm LO}=788\cm$ in pure STO thus provides a direct measure of  $n$.
A distinction between $\mu$ and $d$ is also possible since the former (latter) gives rise to an asymmetric (symmetric) change in the intensity of the maximum and minimum, respectively. 
The result of our analysis is summarized in Fig.~3(c) which displays the data at 10 K together with the best fits using two different models. The green line shows the case of a block-like profile of the electron density on the STO side of the LAO/STO interface which yields $N_{\rm s}=5\times10^{13}\cmtwo$, $d=12$~nm and 
$\mu=10\ {\rm cm^2/Vs}$. While this model does not fully account for the spectra, it provides a clear indication that the electronic charge is spread over a significant range of at least 10~nm. 
The differences with respect to the data, in particular, the missing tail towards higher frequency are a clear indication that $n$ exhibits a sizeable variation along the normal direction. As shown by the red line, an excellent agreement with the experiment is obtained with a model that allows for a depth variation of $n$ (assuming that $n$ is maximal at the LAO/STO interface). The stability of the fitting procedure is outlined in the online material~\cite{EPAPS}.  The corresponding density profile is displayed in Fig.~3(d). It has a sharp maximum at the interface with a maximal carrier concentration of $1.3\times10^{20}\cmthree$ and a full width at half maximum of only 2~nm. In addition, the profile has a pronounced tail that extends to about 11~nm and contains nearly 2/3 of the total sheet carrier concentration. It amounts to $N_{\rm s}^{\rm opt}=5.4\times10^{13}\cmtwo$ (assuming $m^*=3.2 m_e$) and $\mu^{\rm opt}=34\ {\rm cm^2/Vs}$. Notably, the former value agrees well with the one as derived from the Drude response. This is especially true since additional weight may be contained in an extension of the tail beyond 11 nm that is beyond the resolution at this frequency range. Irrespective of this uncertainty, our IR- and transport data consistently yield a value of 
$N_{\rm s}\approx 5-9\times 10^{13}\cmtwo$ that is several times smaller than the prediction of 
$N_{\rm s}=3.3 \times 10^{14}\cmtwo$ of the polarization catastrophe scenario. Assuming this scenario is correct, this implies that the majority of the transferred electrons are strongly trapped with a binding energy, $E_b$, in excess of 0.1 eV. 
Finally, we mention that only a tiny signature of the Berreman mode could be observed in LS-3 [see Fig.~3(c)]. In LS-50 and LS-50-ov the mode is lacking a well resolved maximum in agreement with the much lower $\mu$ values as derived from the Drude response. 

The obtained charge density profile can be well reconciled with previous reports. It is surely consistent with the observation of superconductivity since the distribution of $n$ matches the region of the maximum of the superconducting dome of bulk STO~\cite{Koonce}. Therefore, they agree with a corresponding estimate of $d<10$~nm~\cite{Reyren} and also with a scaling analysis of the anisotropy of the upper critical field which yields $\leq7$~nm~\cite{Schneider}. 
The strongly asymmetric profile also confirms the calculations in Ref.~\cite{Copie} which take into account the large local electric field and the strongly non-linear polarizibility of STO at the interface. Our full width at half maximum of 2~nm is however smaller than the reported experimental value of 12~nm at 10~K~\cite{Copie}.
Our profile also seems consistent with the claim of Ref.~\cite{Sing} that the carrier density decreases very rapidly within the first 4~nm, albeit their measurements were preformed at 300~K where the profile is reported to be narrower~\cite{Copie}. In this comparison one should consider that different definitions of $d$ were used and that the various experimental techniques are either more sensitive to the high concentration at the interface or rather to the long tail which contains most of the weight. In this context we emphasize the significance of our optical data which provide a truly macroscopic (in the lateral direction) and fairly direct probe of the density profile.

Finally, we comment on the marked difference between $\mu^{\rm opt}=34\ {\rm cm^2/Vs}$ as derived from the Berreman mode at $900\cm$ and the corresponding $\mu^{\rm opt}=700\ {\rm cm^2/Vs}$ from the Drude response at low frequency. Notably, a similar frequency dependence of the mobility, or a strong inelastic contribution to the scattering rate, was observed in bulk SrTi$_{1-x}$Nb$_x$O$_3$~\cite{Mechelen}  and explained in terms of polaronic correlations. This suggests that the confined electrons near the LAO/STO interface are also subject to polaronic correlations. 

In summary, with IR ellipsometry and transport measurements we determined the sheet carrier density of $N_{\rm s}\approx 5-9\times10^{13}\cmtwo$ of the electrons at the LAO/STO interface and showed that the additional electrons expected within the electronic reconstruction scenario must be strongly bound with $E_b\geq0.1$~eV. We deduced an effective mass of $m^*\approx 3.2\ m_e$ and a strongly frequency dependent mobility which are similar as in bulk SrTi$_{1-x}$Nb$_x$O$_3$ and thus suggestive of polaronic correlations of the confined electrons. We also determined the vertical profile of the carrier density which is strongly asymmetric with an initial fast decay over 2~nm followed by a pronounced tail that extends to about 11~nm. These results highlight the unique potential of the IR ellipsometry technique to provide quantitative information about the interfacial electronic states in oxide heterostructures.

\begin{acknowledgments}
We benefited from discussions with J. Huml\'\i\v{c}ek, A.J. Millis and the technical  support of Y.-L. Mathis at the ANKA synchrotron at FZ Karlsruhe, D, where a part of the experiments was performed. We acknowledge financial support in Fribourg by the SNF (grant 200020-119784 and NCCR-MaNEP), in Augsburg by the DFG (SFB484) and the EC (Nanoxide), and at Unit\'e Mixte de Physique CNRS by the French ANR program ``Oxitronics''.
\end{acknowledgments}

\end{document}


\onecolumngrid

\baselineskip=16pt
\begin{center}
{\Large\bf
\boldmath EPAPS Supplementary material:\\[0.3cm]
``Dynamical response and confinement of the electrons at the LaAlO$_3$/SrTiO$_3$ interface'' \unboldmath
}
\end{center}
\vspace{0.5cm}

\twocolumngrid


\baselineskip=12pt

\section{Modeling of infrared ellipsometry data} 
Ellipsometry directly measures the tilt angle, $\Psi$, and the eccentricity, $\Delta$, of the polarization ellipse of the reflected light from a sample~\cite{Azzam}. In a case of a semi-infinite sample, the real and imaginary parts of the dielectric function, $\er$ and $\ei$, can be directly obtained from these quantities. In the case of an unknown layer in a heterostructure, the complex dielectric function can be obtained from fitting an appropriate model of the layered structure to the measured $\Psi$ and $\Delta$. 

We have used the standard model of coherent interferences in a layered system~\cite{Azzam} that consists of a LAO layer, a conducting (doped) STO layer, and an insulating semi-infinite STO substrate. A sketch of the layer sequence is shown in Fig. 1. The dielectric function of insulating STO was measured directly on a bare substrate. The dielectric function of LAO was obtained from measurements on a bare LAO substrate that was fitted with Lorentz oscillators whose parameters were adjusted to account for strain effects in the thin LAO layer as to fit the corresponding structures in our experimental data. The dielectric function of the conducting layer of STO was composed of the contributions of a bare STO substrate and a Drude term, $-\omega_{\rm pl}^2/[\omega(\omega+{\rm i}\gamma)]$, where $\gamma$ is the broadening, 
$\omega_{\rm pl}^2=ne^2/(\epsilon_0m^*)$ is the square of the plasma frequency, $n$ is the concentration, $e$ the charge and $m^*$ the effective mass of the electrons. This model was used for the analysis of the far infrared response (Fig. 1 in the paper) in which the parameters of the Drude term were obtained by least squares fitting of the real- and imaginary parts of the pseudo dielectric function. 

\begin{figure}[b]
\vspace*{-0.6cm}
\includegraphics[width=8cm]{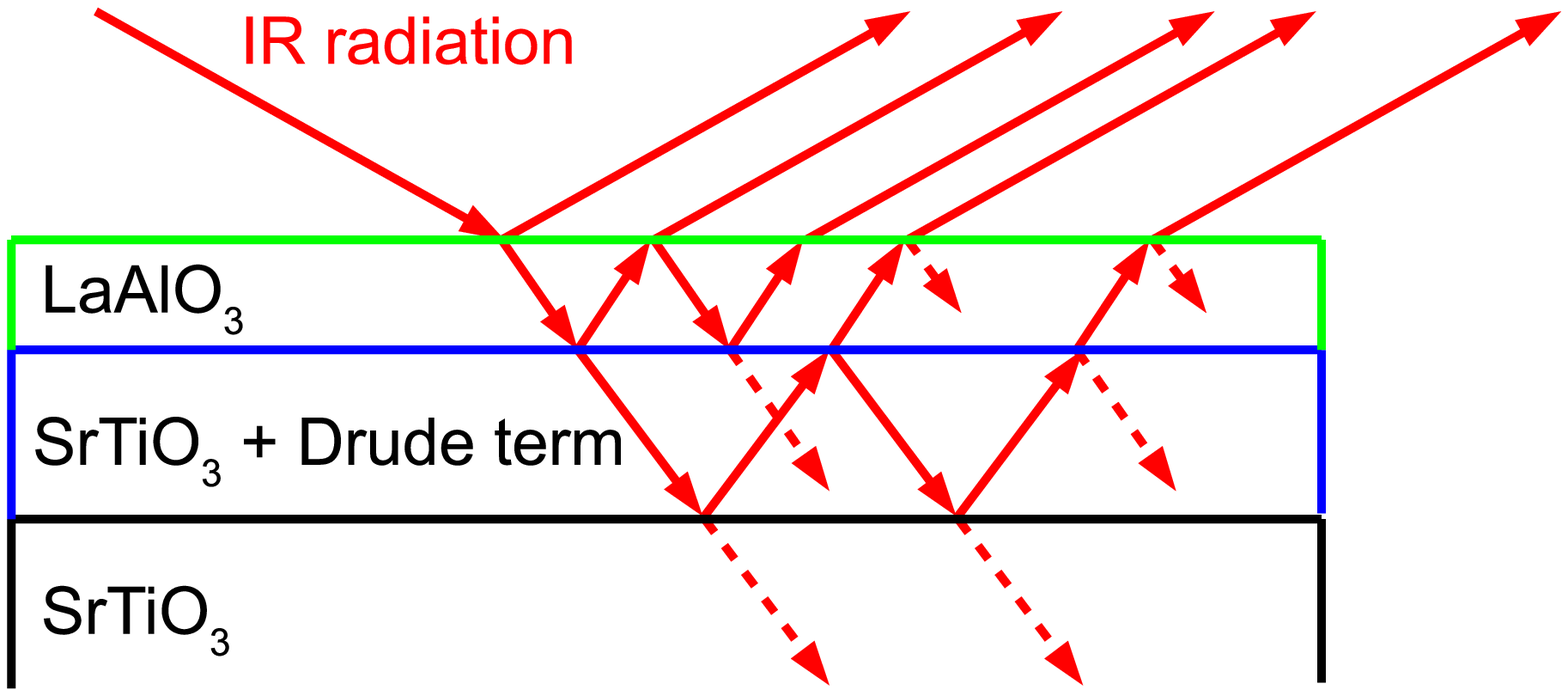}
\vspace*{-0.6cm}
\caption{\label{Model} Sketch of the layer sequence that was used for the modeling of the ellipsometric data.}
\end{figure}

The same model was used for the analysis of the Berreman-mode (Fig. 3 in the paper).  Finally, in the latter case, the oscillator strength of the Drude term was graded across the thickness and fitted to the ellipsometric angle $\Psi$. The stability of the latter fitting procedure is demonstrated by the low value (0.6 nm) of the standard deviation of $d$. In addition, we show in Fig. 2 the evolution of the relative mean square error (MSE) with respect to the best fit value as a function of $d$ which confirms that our solution corresponds to the global minimum. Also shown is a conservative estimate of the extreme boundaries in terms of a 30\% threshold of the MSE increase which yields a range of 8 to 17~nm beyond which $d$ is in essential disagreement with our data.

\begin{figure}[b]
\vspace*{-0.8cm}
\includegraphics[width=7cm]{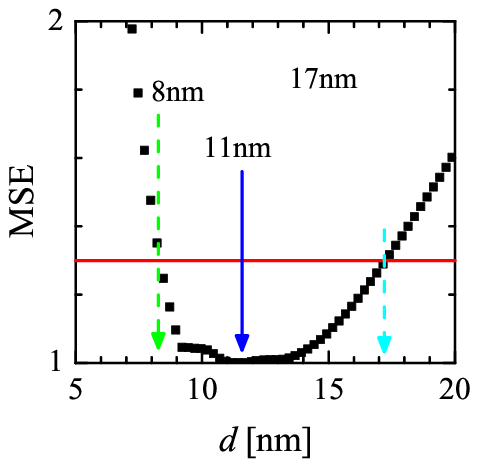}
\vspace*{-0.8cm}
\caption{\label{Model} Relative mean square error (MSE) with respect to the best fit value as a function of $d$. The horizontal line marks a conservative estimate of the boundaries in terms of a 30\% threshold of the MSE increase.  }
\end{figure}

\begin{figure}[h!t]
\vspace*{-0.8cm}
\hspace*{-0.8cm}
\includegraphics[width=10cm]{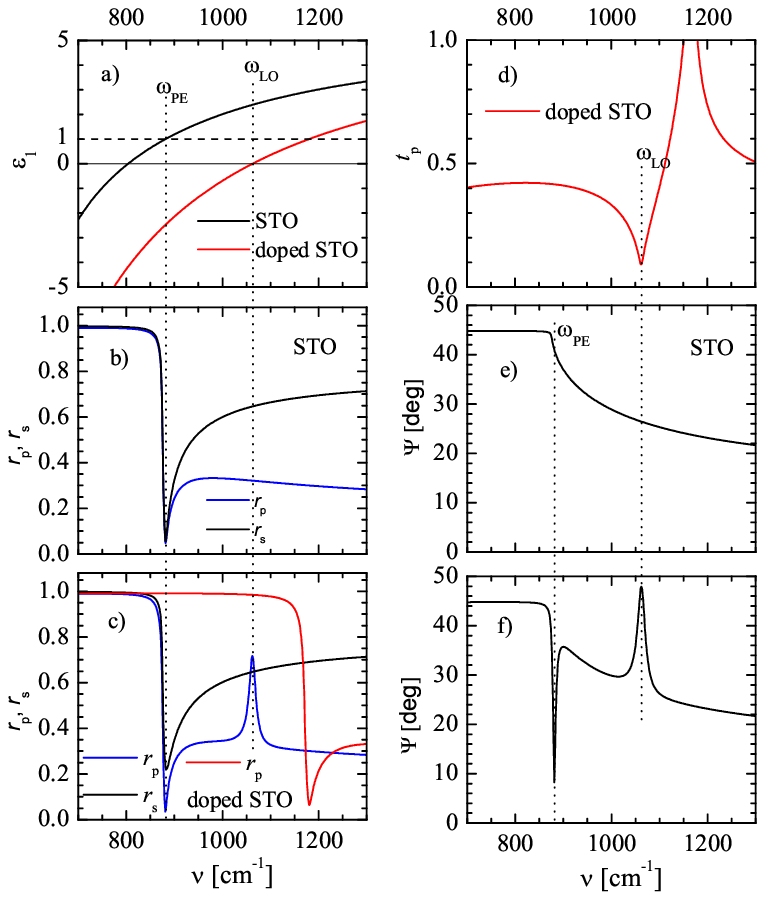}
\vspace*{-0.8cm}
\caption{\label{Model} Depicting the origin of the Berreman mode. (a), Real part of the dielectric functions of bulk samples of undoped and doped STO. (b), Reflection coefficients of an undoped STO substrate. (c) Reflection coefficients of the heterostructure consisting of an undoped STO substrate and a 50~nm thick layer of doped STO on top (black and blue lines) as well as for a doped STO substrate (red line). (d) p-transmission Fresnel coefficient of a doped STO substrate. (e) Ellipsometric angle $\Psi$ for a undoped STO substrate. (f) Ellipsometric angle $\Psi$ for the heterostructure consisting of a STO substrate that is covered with a 50~nm thick layer of conducting STO.  All parameters are calculated for an angle of incidence of 75 degree.  }
\end{figure}

\section{Far-infrared response}
In the paper, we discuss the comparison of the far-infrared experimental spectra and the results of the model described above with the thickness of the conducting layer of 10~nm and block profile of the charge concentration (see middle and lower panels in Fig. 1(b)-(e) of the paper). We show that it well explains the main trends in $\er$ and $\ei$ due to the charge carriers. The model reproduces even minor features in the experimental spectra. The s-shaped resonance near $390\cm$ corresponds to a transverse optical (TO) phonon mode of the strained LAO layer.  The strength of this resonance correlates with the thickness of the LAO layers. It is thus strongest in LS-50 and LS-50-ov, much weaker in LS-5, and hardly resolved in LS-3. The second s-shape resonance near $280\cm$ which is noticeable in the spectra of LS-50 and LS-50-ov arises from a Berreman mode related to an LO frequency of the LAO layer. The sharper feature near $440\cm$ arises from an additional STO phonon mode which is induced by the lattice distortions below the anti-ferrodistortive transition at 104~K~\cite{Petzelt} and the one near $480\cm$ corresponds to a so-called Berreman mode due to the conducting layer related with the LO frequency of STO near $480\cm$. It is discussed in more detail below.

\section{Berreman-modes}
Here we discuss in more detail the origin of the Berreman modes whose spectral features are fully accounted for by the well know formalism of Fresnel coefficients and coherent interferences of radiation in a layered structure. Our aim is to elucidate the origin of the various spectroscopic features and to emphasize which information they can provide about the charge carrier density near the LAO/STO interface.
Figure 3(a) shows the real part of the model dielectric function of insulating (bare) STO (full black line) as well as of doped and thus conducting STO (full red line) around the frequency of the highest longitudinal-optical (LO) phonon mode. The dielectric function of the conducting STO consists of the sum of the response of insulating STO and a Drude-term with $n = 10^{20}\cmthree$ and $\mu=1000\ {\rm cm^2/Vs}$
(we assumed $m^*=3.2 m_e$). 
The almost negligible and featureless absorptive part of the dielectric function is not shown. 
Figure 3(b) shows for insulating STO the corresponding p- and s- Fresnel reflection coefficients,
 $r_{\rm p}$ and $r_{\rm s}$, respectively, for an angle of incidence of 75 degree. The dip near $865\cm$ occurs at the frequency of the so-called plasma edge, $\omega_{\rm PE}$, which fulfills the condition $\epsilon_1(\omega_{\rm PE}) = 1 $ (that matches the value in vacuum). 
Figure 3(c) shows the reflection coefficients (blue and black line) for the case of a STO substrate that is covered with a 50~nm thick layer of conducting STO which is similar to the heterostructures that are discussed in the paper. In these simulations, we have neglected the 2~nm layer of LAO since it has an almost negligible influence on the response functions in this frequency range. For comparison we also show the Fresnel coefficient $r_{\rm p}$ of the doped STO layer (red line) where the plasma edge in $r_{\rm p}$ is shifted to a higher frequency of about $1180\cm$. This upward shift is due to the inductive free carrier response which reduces the total value of $\er$. 
The $r_{\rm p}$ reflection coefficient for the heterostructure (blue line) follows rather closely the one of the bare substrate (blue line in Fig. 3(b) except for the region around the LO-mode of the doped layer with $\er(\omega_{\rm LO}) = 0$ at $1060\cm$ that exhibits a sharp peak, so-called Berreman mode; see 
Ref.~[21] of the paper. This mode arises from a dynamical charge accumulation at the interfaces of the heterostructures that is driven by the normal component of the polarized light. A set of two or more interfaces accordingly gives rise to an IR-active dipole and a corresponding resonance near the frequency of the LO modes of the constituent materials~\cite{Sihvola}. The null point in the Fressnel p-transmission coefficient [see Fig. 3(d)] is related to the singularity.

Figures 3(e) and 3(f) show the ellipsometric angle, $\Psi$, as given by the ratio of the Fresnel coefficients, $\tan(\Psi)= |r_{\rm p}/r_{\rm s}|$, for the insulating STO substrate and the heterostructure, respectively. For the STO substrate $\Psi$ exhibits no marked features besides a change in slope at $\omega_{\rm PE}$. In contrast, for the heterostructure $\Psi$ exhibits a pronounced peak at $\omega_{\rm LO}$ of the conducting layer [similarly to $r_{\rm p}$ in Fig. 3(c)] and a dip at 
$\omega_{\rm PE}$ of the STO substrate. The latter arises because the presence of the thin doped layer gives rise to noticeable differences in the dips of $r_{\rm p}$ and $r_{\rm s}$ around $\omega_{\rm PE}$  [see Fig. 3(c)]. The doped layer in the heterostructure thus gives rise to a dip that is located at $\omega_{\rm PE}$ of the undoped STO substrate and a peak at the $\omega_{\rm LO}$ of the bulk doped STO, respectively.  These are the same dip and peak structures as seen in the experimental data in Fig. 3 of the paper. However, as the position of the peak strongly depends on the carrier density which in a real sample exhibits a graded vertical profile, this peak structure becomes broadened and exhibits a tail towards higher frequency.

Finally we note, that similar dip features due to additional Berreman-modes at the LO-mode of STO at 
$482\cm$ and the LO-mode of LAO at $280\cm$ appear in the experimental spectra as shown in Figs.~1(b) and 1(e) of the paper, respectively. In both cases the peaks turn out to be extremely weak and close to the noise level of the experiment. This is mostly due to the circumstance that the real part of the dielectric function rises here very steeply due to the TO-mode of another IR-active phonon which is located at a slightly higher frequency. For this reason we found that it is much more difficult to deduce from these Berreman-modes reliable values for the thickness of the conducting layer.